\documentclass[prb,twocolumn,superscriptaddress,floatfix,showpacs]{revtex4}
\usepackage{graphicx,amsfonts,amssymb,amsmath,hyperref,enumerate}

\newif\ifhyper
\hypertrue
\ifhyper
\hypersetup{
   citecolor = {green},
   colorlinks = {true}, 
   urlcolor = {blue} 
}
\fi

\newcommand{\beq}{\begin{equation}}
\newcommand{\eeq}{\end{equation}}
\newcommand{\beqa}{\begin{eqnarray}}
\newcommand{\eeqa}{\end{eqnarray}}

\def\Longarrow{\protect\@lra}
\def\@lra{\relbar\joinrel\relbar\joinrel\relbar\joinrel%
          \relbar\joinrel\rightarrow}

\begin{document}


\title{All spin-1 topological phases in a single spin-2 chain}

\author{Augustine Kshetrimayum}
\affiliation{Institute of Physics, Johannes Gutenberg University, 55099 Mainz, Germany}

\author{Hong-Hao Tu}
\affiliation{Max-Planck-Institut f\"ur Quantenoptik, Hans-Kopfermann-Str.~1, 85748 Garching, Germany}

\author{Rom\'an Or\'us}
\affiliation{Institute of Physics, Johannes Gutenberg University, 55099 Mainz, Germany}

\begin{abstract}

Here we study the emergence of different Symmetry-Protected Topological (SPT) phases in a spin-2 quantum chain. We consider a Heisenberg-like model with bilinear, biquadratic, bicubic, and biquartic nearest-neighbor interactions, as well as uniaxial anisotropy. We show that this model contains  four different effective spin-1 SPT phases, corresponding to different representations of the $(\mathbb{Z}_2 \times \mathbb{Z}_2) + T$ symmetry group, where $\mathbb{Z}_2$ is some $\pi$-rotation in the spin internal space and $T$ is time-reversal. One of these phases is equivalent to the usual spin-1 Haldane phase, while the other three are different but also typical of spin-1 systems. The model also exhibits an $SO(5)$-Haldane phase. Moreover, we also find that the transitions between the different effective spin-1 SPT phases are continuous, and can be described by a $c=2$ conformal field theory. At such transitions, indirect evidence suggests a possible effective field theory of four massless Majorana fermions. The results are obtained by approximating the ground state of the system in the thermodynamic limit using Matrix Product States via the infinite Time Evolving Block Decimation method, as well as by effective field theory considerations. Our findings show, for the first time, that different large effective spin-1 SPT phases separated by continuous quantum phase transitions can be stabilized in a simple quantum spin chain.

\end{abstract}

\pacs{75.10.Pq, 75.10.Jm}
\maketitle

\section{Introduction} Topological order \cite{to} is a new kind of order in quantum matter. Such order can be protected by certain symmetries, i.e., it is present unless the symmetries are broken. This is the concept of Symmetry-Protected Topological (SPT) phases, discussed originally in the Haldane phase of the spin-1 quantum Heisenberg chain \cite{aklt}, but relevant to higher-dimensional systems as well \cite{weyl}. SPT phases had a recent revival thanks to concepts such as entanglement spectrum \cite{es}. In particular, using the language of Matrix Product States and Tensor Networks \cite{tn} it was realised that $1d$ SPT phases are related to degeneracy patterns in the eigenvalue spectrum of reduced density matrices of the chain \cite{poll1}. Other approaches based on MPS \cite{param0, param, ndi} and group theory \cite{wen1, wen2} have also been successful in characterising $1d$ SPT phases.

In this context, Oshikawa conjectured in 1992 that the spin-2 quantum Heisenberg chain with uniaxial anisotropy should have an effective spin-1 SPT phase similar to the usual Haldane phase, commonly called an Intermediate-Haldane (IH) phase \cite{Oshik}. Such an effective spin-1 phase remained elusive for many years, and could only be found recently \cite{IHall}. Yet, the relative size of this phase is quite small in parameter space, which easily makes it fragile against noise. A different approach was taken in Ref.~\cite{TuOrus}, where a generalized spin-2 Heisenberg chain was considered with bilinear, biquadratic, bicubic and biquartic interactions. Such a model can be mapped, in a specific regime of parameters, to an $SO(5)$-symmetric model which, in the presence of uniaxial anysotropy, has (i) a very large effective spin-1 IH phase, and (ii) a small ``$SO(5)$-Haldane" phase which is also SPT. This construction, however, relied on a precise fine-tuning of the parameters in the system.

The intermediate SPT phase found in all the above models is equivalent to the well-known spin-1 Haldane phase \cite{aklt}. However, this is not the only SPT phase realizable for spin-1 \cite{wen2, wen3}, and hence this should not be the only possibility to emerge as an effective spin-1 phase of a spin-2 quantum chain. Still, it is  rather difficult to find simple and realistic examples of quantum spin chains with different (perhaps effective) spin-1 SPT phases and, even more difficult, where these phases are separated by continuous quantum phase transitions.

In this paper we solve the above problems by studying a spin-2 Heisenberg-like model with arbitrary values of bilinear, biquadratic, bicubic and biquartic nearest-neighbor interactions, together with a uniaxial anisotropy. For this model we show that different effective spin-1 SPT phases can actually be stabilized and, moreover, that these are separated by quantum critical points. To be specific, our different SPT phases correspond to different representations of the $(\mathbb{Z}_2 \times \mathbb{Z}_2) + T$ symmetry group, where $\mathbb{Z}_2$ is some $\pi$-rotation in the spin internal space and $T$ is time-reversal. One of these phases, called $T_0$ \cite{wen2}, is equivalent to the usual spin-1 Haldane phase. We find that the transitions between these phases can be described by a $c=2$ conformal field theory (CFT), and indirect evidence suggests that this may be related to an effective field theory of four massless Majorana fermions. The model also exhibits other features, such as the $SO(5)$-Haldane phase. The results are obtained by approximating the ground state of the system in the thermodynamic limit using Matrix Product States \cite{tn} via the infinite Time Evolving Block Decimation method \cite{itebd}, as well as by an effective field theory description \cite{effective, neweff}. Our findings show, for the first time, that different effective spin-1 SPT phases separated by quantum critical points can \emph{de facto} Êbe stabilized in a simple spin-2 Heisenberg-like quantum spin chain.

The paper is organized as follows: in Sec. II we present the model and its symmetries. In Sec. III we discuss our approach to study the model. Sec. IV includes our results for  different projections of the phase diagram. In Sec. V we discuss a effective field theory approach in terms of Majorana fermions. In Sec. VI we comment briefly on other features observed in the phase diagram of the model. Finally, Sec. VII includes our conclusions. In the appendices we discuss in some detail the 16 SPT phases in $1d$ protected by $(\mathbb{Z}_2 \times \mathbb{Z}_2) + T$ symmetry, and how to extract some "SPT order parameters" from the Matrix Product State that approximates the ground state of the system. 

\begin{figure}
\includegraphics[width=0.45\textwidth]{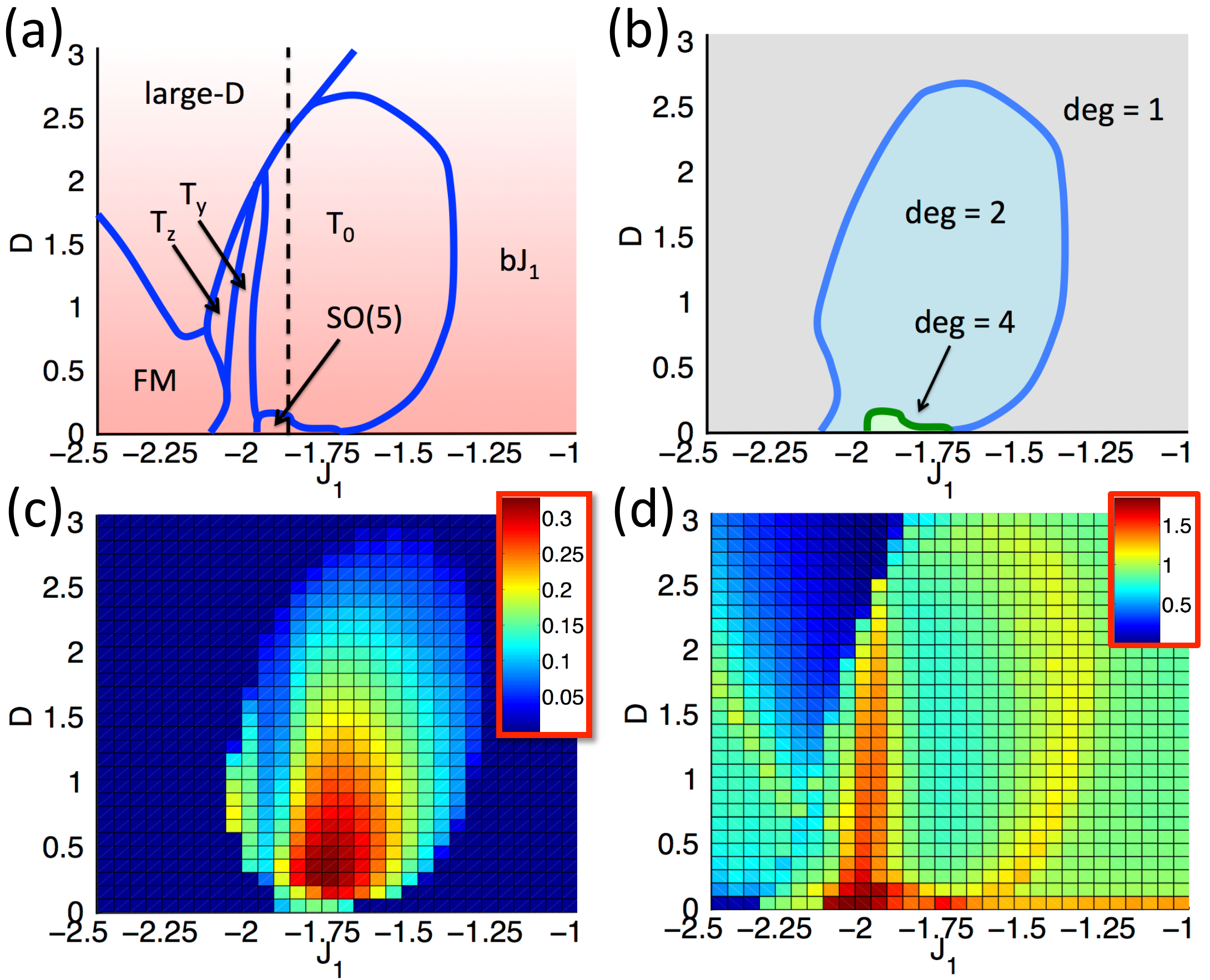}
\caption{(a) Phase diagram for the $\langle J_1,D \rangle$ plane. The dotted line is the one studied in Ref.~\cite{TuOrus}; (b) degeneracies in the entanglement spectrum; (c) string-order parameter $O^{34}$ -- aerial view -- ; (d) entanglement entropy of half an infinite chain -- aerial view --. }
\label{FigJ1}
\end{figure}
\begin{figure}
\includegraphics[width=0.43\textwidth]{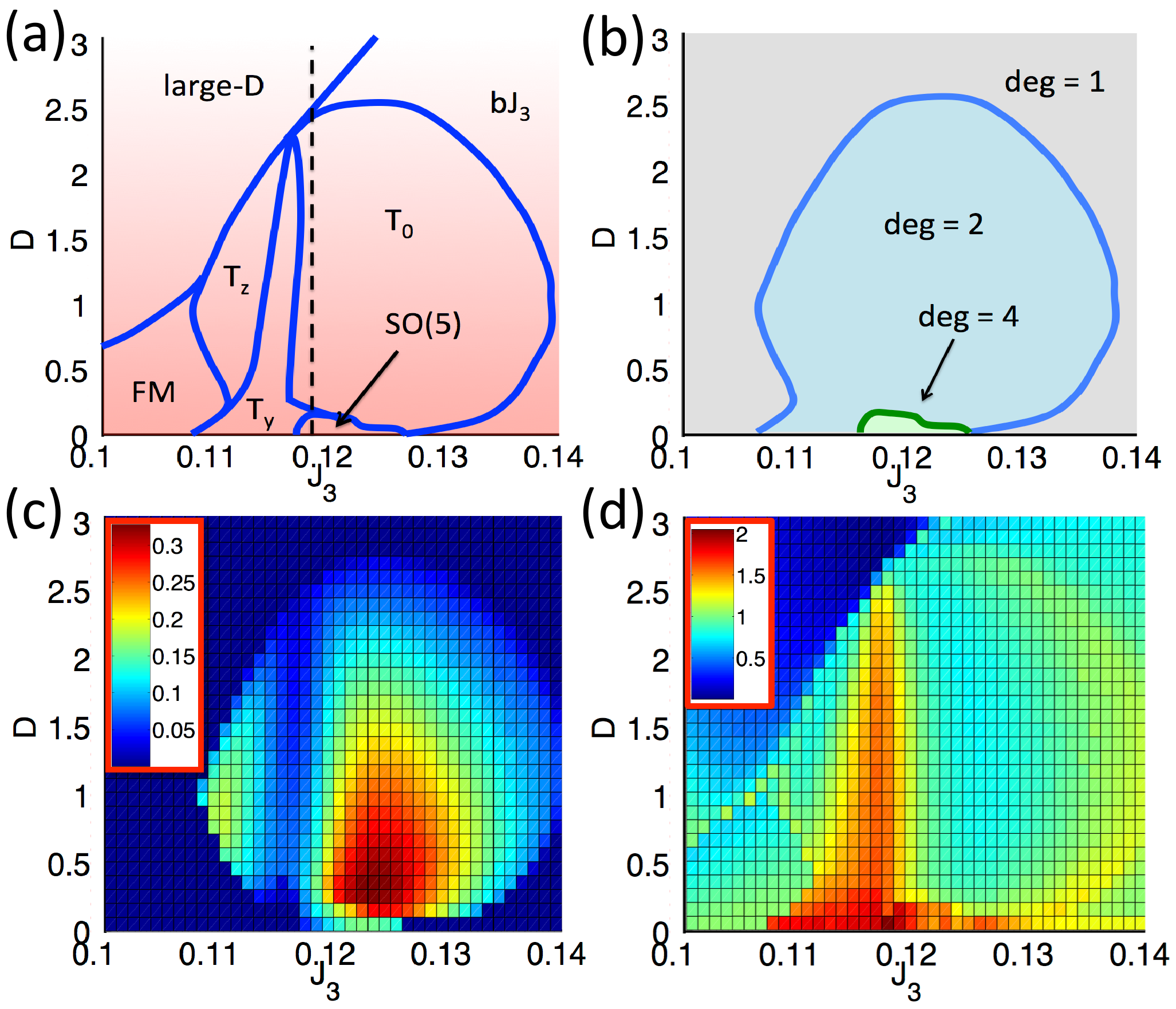}
\caption{(a) Phase diagram for the $\langle J_3,D \rangle$ plane. The dotted line is the one studied in Ref.~\cite{TuOrus}; (b) degeneracies in the entanglement spectrum; (c) string-order parameter $O^{34}$ -- aerial view -- ; (d) entanglement entropy of half an infinite chain -- aerial view --. }
\label{FigJ3}
\end{figure}

\section{Model and symmetries} Here we consider the spin-2 quantum chain
\begin{equation}
H=\sum_{j}\sum_{\gamma =1}^{4}J_{\gamma }(\vec{S}_{j}\cdot \vec{S}%
_{j+1})^{\gamma }+D\sum_{j}(S_{j}^{z})^{2},  \label{eq:Hamiltonian}
\end{equation}%
 for periodic boundary conditions and in the thermodynamic limit. In a certain regime of parameters, the model for $D=0$ is also known to have ferromagnetic,  dimerized, and critical trimerized phases \cite{pochung}. In Ref.~\cite{TuOrus}, the case $J_{1}=-\frac{11}{6},J_{2}=-\frac{31}{180},J_{3}=\frac{11}{90},J_{4}=\frac{1}{60}$ was addressed also for $D\geq 0$. In this regime it was proven that, for $D=0$, the system has an exact $SO(5)$ symmetry and an MPS as its
exact ground state \cite{Tu-2008, Scalapino-1998}. For $D>0$ the $SO(5)$ symmetry is explicitly broken down to $U(1)\times U(1)$ \cite{TuOrus}. Moreover, for these values of the parameters, the Hamiltonian in Eq.~(\ref{eq:Hamiltonian}) also has discrete symmetries,
including spatial inversion $P$, time reversal $T$, and a $(\mathbb{Z}_2\times \mathbb{Z}_2)^{2}$ related to invariance under global $\mathbb{Z}_2$ rotations. As discussed in Ref.~\cite{TuOrus}, these symmetries protect both an IH phase and the $SO(5)$-Haldane phase. Phase transitions in this system were also studied using an effective field theory of five Majorana fermions \cite{effective, TuOrus, neweff}.

If we also allow for a change in the values of $J_\gamma$, then the Hamiltonian above has a $(U(1) \times \mathbb{Z}_2) + T$ symmetry, where $U(1)$ corresponds to the $S_z$ conservation. We focus, however, on the reduced discrete symmetry $(\mathbb{Z}_2 \times \mathbb{Z}_2) + T$, so that in principle we could also add terms breaking the $U(1)$ symmetry down to $\mathbb{Z}_2$. Such a symmetry is known to protect up to $16$ different possible SPT phases \cite{wen1, wen2}. Four of these phases are typical of spin-1 chains, and following the notation in Ref.~\cite{wen2} we call them $T_0, T_x, T_y$ and $T_z$, with $T_0$ the usual Haldane phase for spin-1 chains \cite{aklt}.

\section{Approach} In this paper we study the phase diagram of the above model for arbitrary values of the interaction strengths $J_\gamma$ and anisotropy $D$, thus the symmetry $(\mathbb{Z}_2 \times \mathbb{Z}_2) + T$ turns out to be relevant for us. For the sake of simplicity, we focus on four two-dimensional projections of the phase diagram  obtained by fixing all the $J_\gamma$ except one to the values in Ref.~\cite{TuOrus} (i.e., the ones mentioned above). Thus, we study the four two-dimensional planes $\langle J_1,D \rangle$,$\langle J_2,D \rangle$, $\langle J_3,D \rangle$ and $\langle J_4,D \rangle$, with the rest of interaction parameters fixed to the values mentioned for each case.

\begin{figure}
\includegraphics[width=0.41\textwidth]{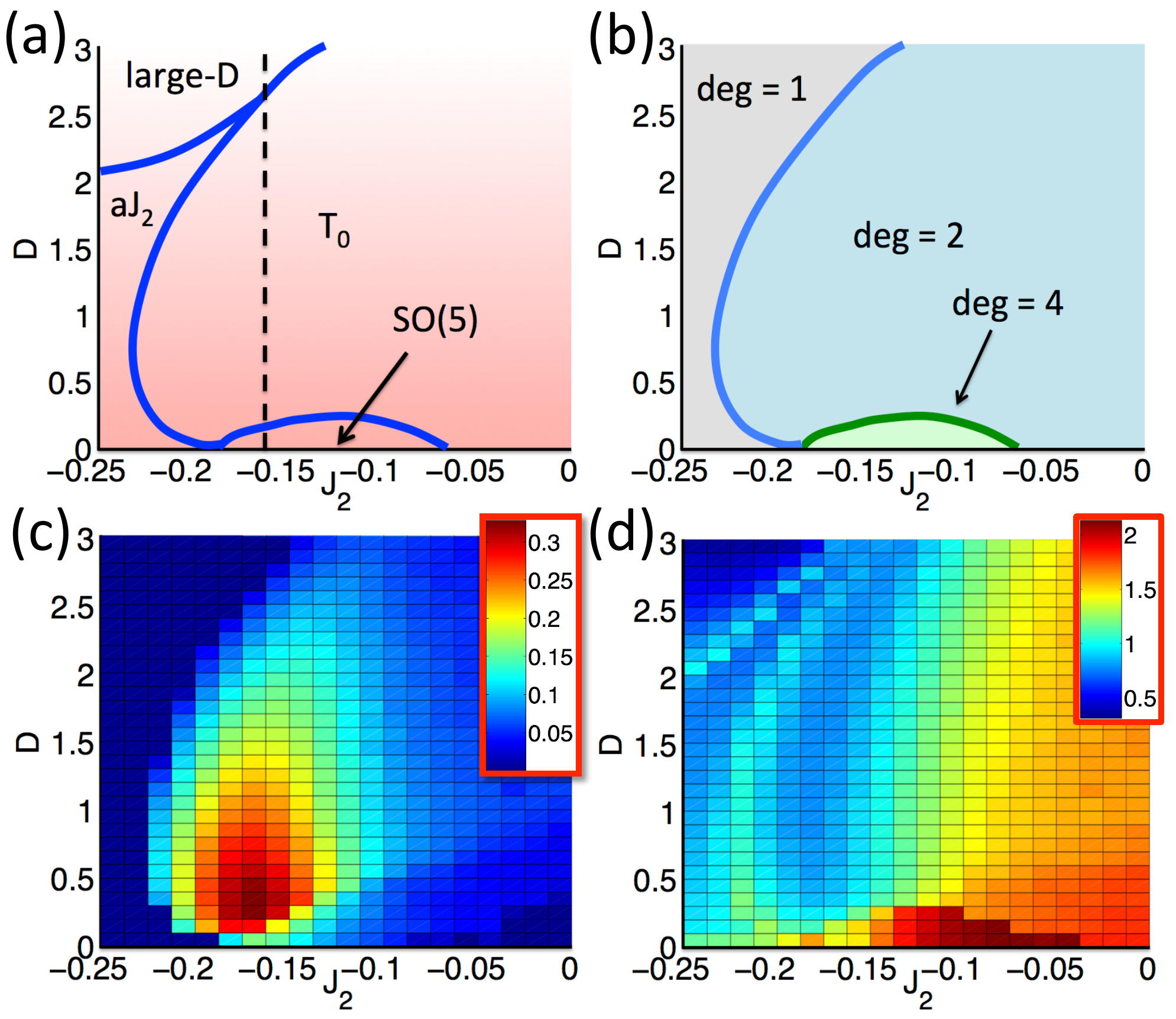}
\caption{(a) Phase diagram for the $\langle J_2,D \rangle$ plane. The dotted line is the one studied in Ref.\cite{TuOrus}; (b) degeneracies in the entanglement spectrum; (c) string-order parameter $O^{34}$ -- aerial view -- ; (d) entanglement entropy of half an infinite chain -- aerial view --. }
\label{FigJ2}
\end{figure}

We approximate the ground state of the system in the thermodynamic limit by a Matrix Product State (MPS) using the infinite Time-Evolving Block Decimation method \cite{itebd}. The maximum MPS bond dimension is around $120$, which proves sufficient for our purposes. For each plane, we evaluate the expectation value of string order parameters
\begin{equation}
\mathcal{O}^{12}=\lim_{|k-j|\rightarrow \infty }\langle
L_{j}^{12}\prod_{l=j+1}^{k-1}\exp (i\pi L_{l}^{12})L_{k}^{12}\rangle
\label{eq:SOP}
\end{equation}%
and $\mathcal{O}^{34}$ (where $L^{12}$ is replaced by $L^{34}$), with $L^{12}=|2\rangle \langle
2|-|-2\rangle \langle -2|$ and $L^{34}=|1\rangle \langle 1|-|-1\rangle
\langle -1|$ in the basis of spin-2. The string order parameters $\mathcal{O}^{34}$ and $\mathcal{O}^{12}$ measure the hidden antiferromagnetic order in $|\pm 2\rangle$ and $|\pm 1\rangle$ sectors, respectively. When close to quantum critical points in the Ising universality class (described by a Majorana-fermion effective theory, as we shall see in Eq. (\ref{eq:FT})), presence/absence of string orders indicate different sorts of Ising ordered/disordered gapped phases \cite{effective} (reflecting the signs of Majorana masses in Eq.~(\ref{eq:FT})), thus providing a partial characterization of the gapped phases and also a hint for the underlying CFT. Moreover, we also compute the degeneracies in the entanglement spectrum and the entanglement entropy of half an infinite chain (within the limitations of our finite bond dimension). All this allows us to see the potential candidates for SPT phases in the model. To determine exactly which type of SPT phases we have, we compute the parameters $(\beta, \omega, \mu, \nu)$ \footnote{See appendix.} defined as
\beqa
R_t^2 &=& \beta \mathbb{I}  \\
R_xR_z &=& \omega R_z R_x  \\
R_xR_t &=& \mu R_t R_x  \\
R_zR_t &=& \nu R_t R_z,
\eeqa
with $R_t$, $R_x$ and $R_z$ the matrix representations of, respectively, time reversal $T$, the $\pi$-rotation around the $x$-axis, and the $\pi$-rotation around the $z$-axis, acting on the MPS bond indices (with the convention $R_x^2 = R_z^2 = \mathbb{I}$). It turns out that such matrices can be computed easily using MPS techniques \cite{param0, param}. The numbers $(\beta, \omega, \mu, \nu)$ are all equal to $\pm 1$, and their 16 different choices correspond to the 16 SPT phases protected by $(\mathbb{Z}_2 \times \mathbb{Z}_2) + T$ symmetry, in one-to-one match with those in Tab. I of Ref.~\cite{wen2}. In the appendix we review this classification, as well as how to find $(\beta, \omega, \mu, \nu)$ for an MPS.

\begin{figure}
\includegraphics[width=0.43\textwidth]{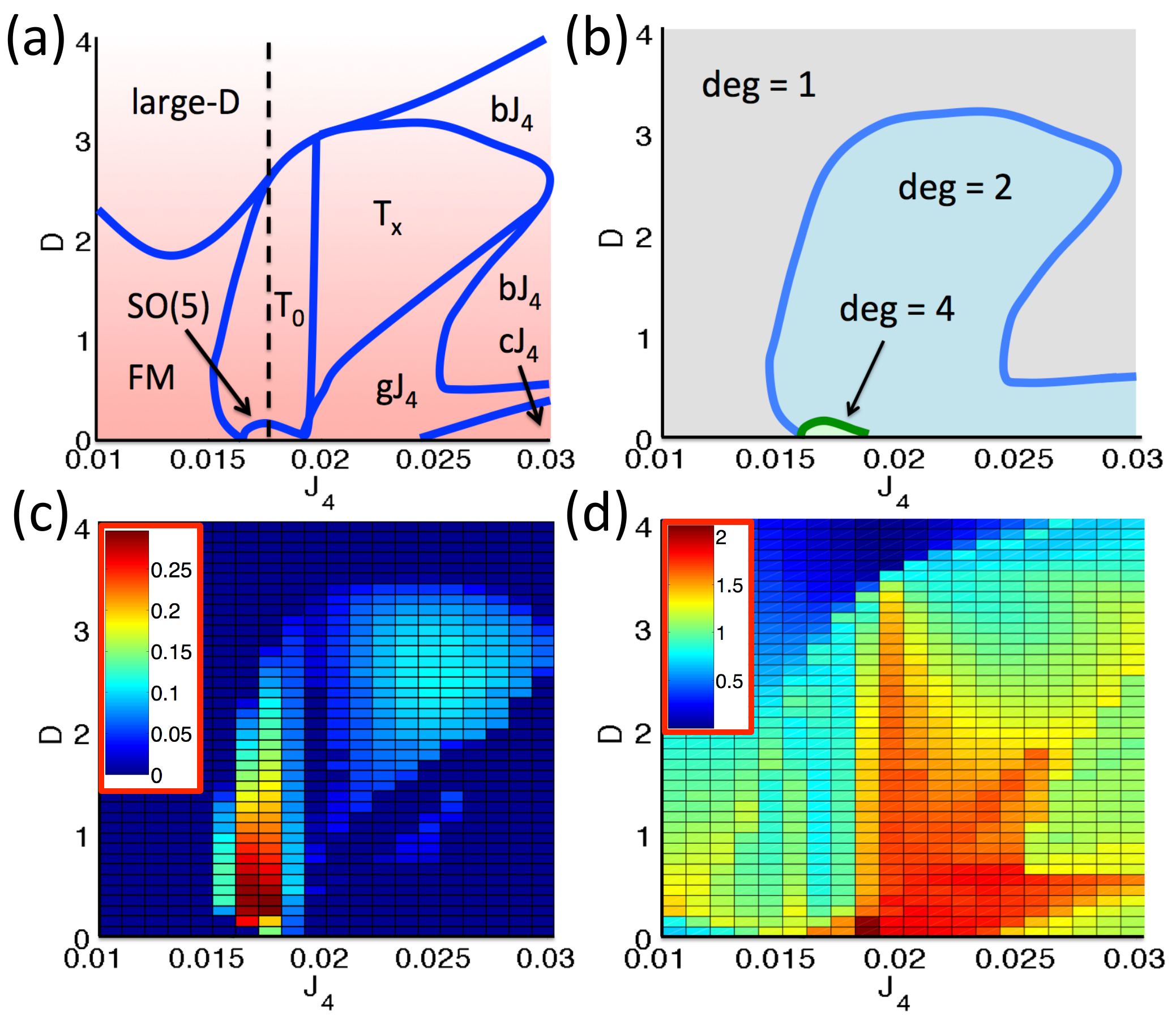}
\caption{(a) Phase diagram for the $\langle J_4,D \rangle$ plane. The dotted line is the one studied in Ref.\cite{TuOrus}; (b) degeneracies in the entanglement spectrum; (c) string-order parameter $O^{34}$ -- aerial view -- ; (d) entanglement entropy of half an infinite chain -- aerial view --. }
\label{FigJ4}
\end{figure}

\section{Phase diagram projections} 
\begin{figure}
\includegraphics[width=0.45\textwidth]{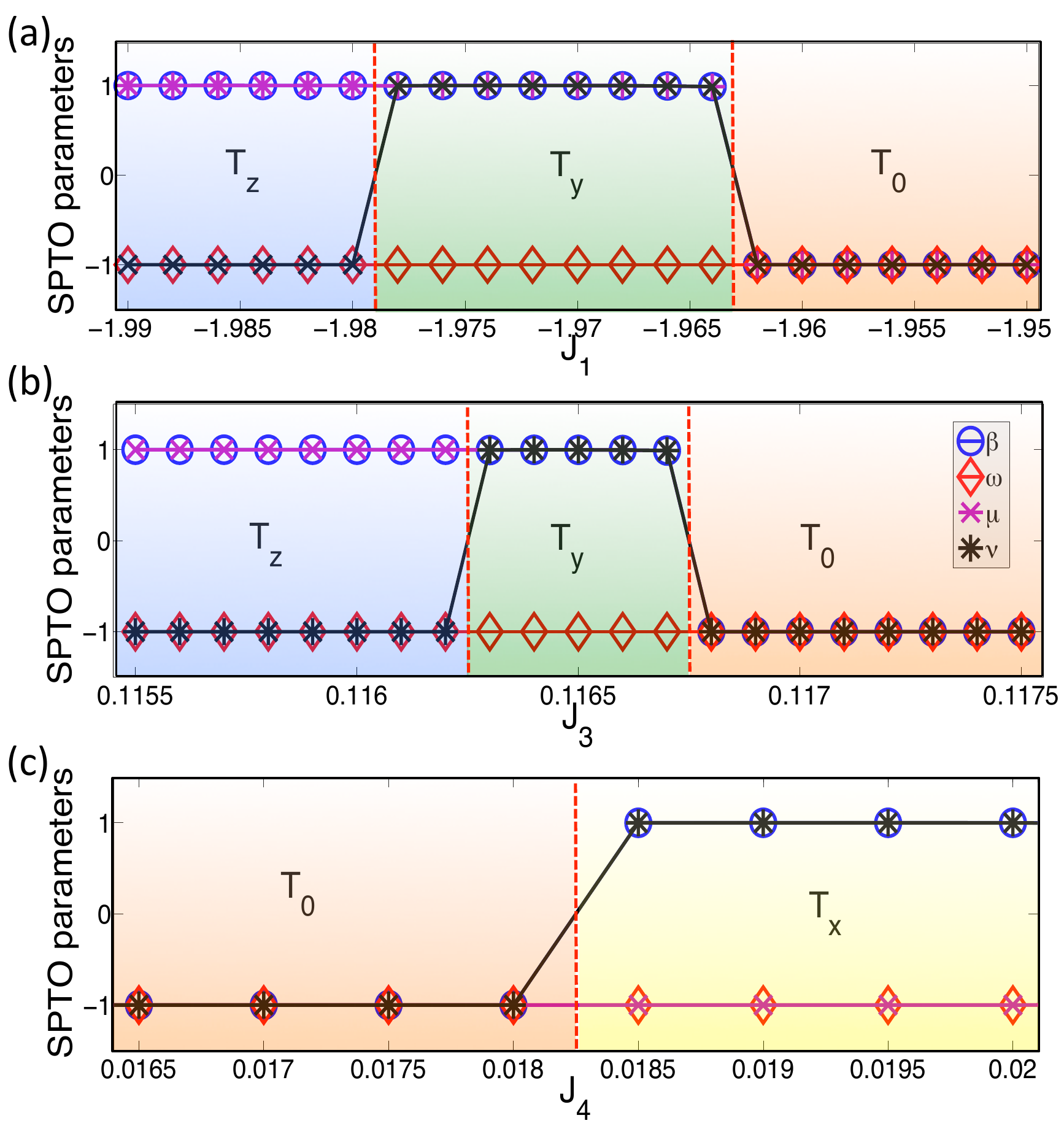}
\caption{SPT parameters $(\beta, \omega, \mu, \nu)$ for (a) the $\langle J_1,D \rangle$ plane at $D=1$, (b) the $\langle J_3,D \rangle$ plane at $D=1.5$, and (c) the $\langle J_4,D \rangle$ plane at $D=1$. The data are consistent with four different effective spin-1 SPT phases, separated by the vertical red dashed lines.}
\label{FigOrders}
\end{figure}

Our results for the phase diagrams can be found in Figs.~\ref{FigJ1}-\ref{FigJ4}. In all cases, we find a relatively small $SO(5)$-phase with degeneracy $4$ in the entanglement spectrum and non-zero value of $O^{12}$ (not shown), and a quite large region with degeneracy $2$ in the entanglement spectrum, as well as non-zero values of $O^{34}$, compatible in principle with a large IH ($T_0$) phase. We see also a strong similarity between the diagrams in the $\langle J_1,D \rangle$ and $\langle J_3,D \rangle$ planes, see Figs.~\ref{FigJ1}-\ref{FigJ3}. In particular, our results for each of these two  diagrams are compatible with \emph{three different intermediate spin-1 SPT phases}, see Fig.~\ref{FigOrders}(a,b), namely, $T_z, T_y$ and $T_0$. Surprisingly, we find that the $T_y$ phase has a very large entanglement entropy. There are two plausible scenarios to explain this: either (i) it is a gapped phase with very small gap, or (ii) it is a gapless critical phase. We have explicitly checked that the entanglement entropy seems to increase when increasing the bond dimension in this phase, and therefore we believe that scenario (ii) is more plausible. If this is indeed the case, the $T_y$ phase in our model exhibits both gapless degrees of freedom and, to some extent, SPT order, which deserves further investigations \footnote{There are examples of gapless SPT phases, see e.g., Ref. \cite{weyl}. But their classification is still far from clear.}. While for the $\langle J_2,D \rangle$ diagram we find only one candidate for an intermediate topological phase ($T_0$), in the $\langle J_4,D \rangle$ plane we find two: $T_0$ and $T_x$, see Fig.~\ref{FigOrders}(c). In Tab.~\ref{tabd} we summarize all the effective spin-1 phases found so far.

\begin{table}
\begin{center}
\begin{tabular}{||c||c||c||c||c||c||}
\hline
~~~Phase~~~  &  ~~~~Figure~~~~  & ~~$ \beta $~~ & ~~$ \omega $~~ & ~~$ \mu $~~ & ~~$ \nu $~ \\
 \hline
 \hline
$T_0 = $ IH & Figs.\ref{FigJ1},\ref{FigJ3},\ref{FigJ2},\ref{FigJ4} & -1 & -1 & -1 & -1  \\
\hline
$T_x$ & Fig.\ref{FigJ4} & +1 & -1 & -1 & +1  \\
\hline
$T_y$ & Figs.\ref{FigJ1},\ref{FigJ3} & +1 & -1 & +1 & +1  \\
\hline
$T_z$ & Figs.\ref{FigJ1},\ref{FigJ3} & +1 & -1 & +1 & -1  \\
\hline
\end{tabular}
\end{center}
\caption{Different effective spin-1 phases found in Figs.~\ref{FigJ1}-\ref{FigJ4}, protected by $(\mathbb{Z}_2\times \mathbb{Z}_2) + T$ symmetry.}
\label{tabd}
\end{table}
\begin{figure}
\includegraphics[width=0.43\textwidth]{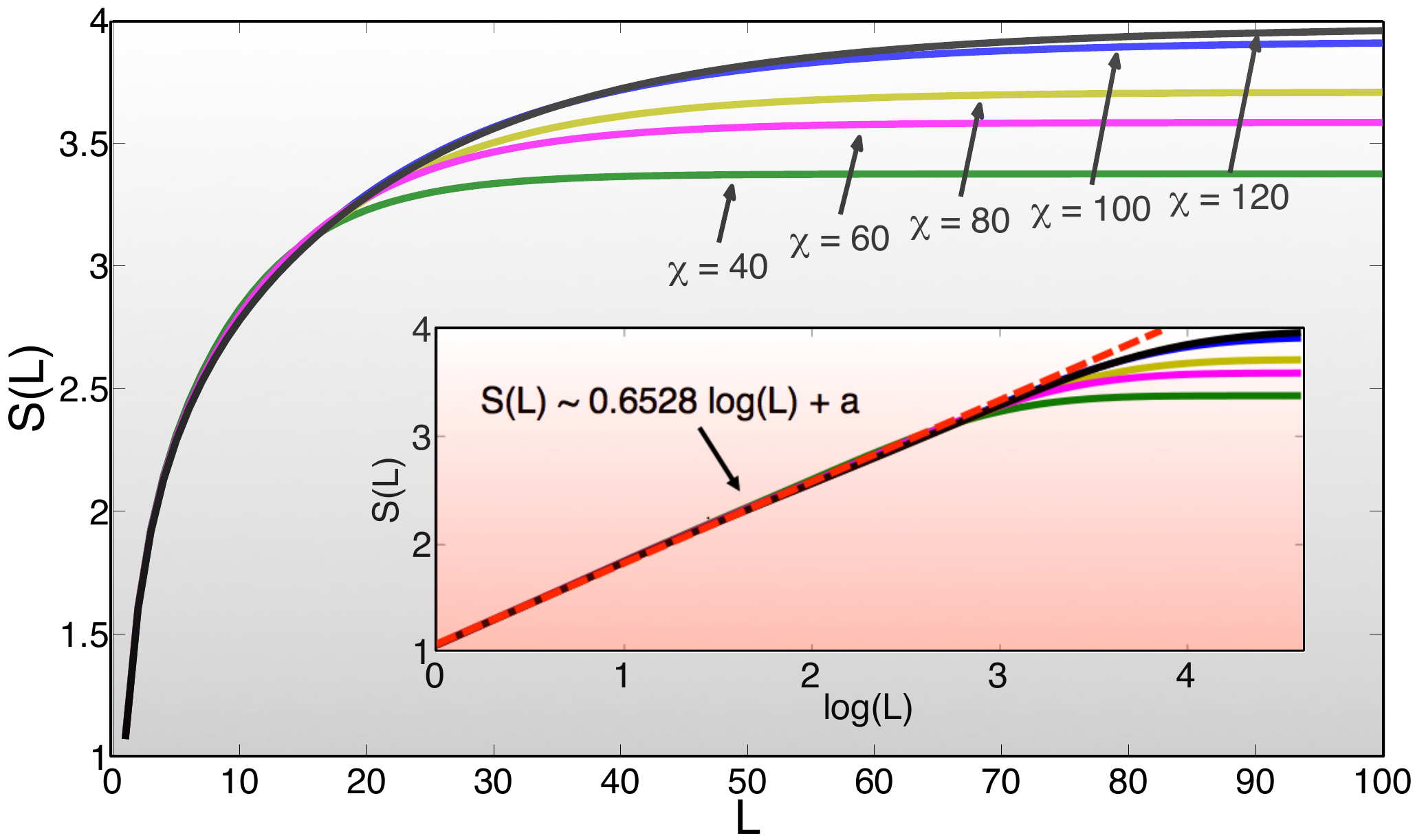}
\caption{Scaling of the entanglement entropy $S(L)$ of a block of length $L$ for the $T_y-T_0$ transition in Fig.\ref{FigJ3} at $D=1.5$. In the inset we plot the same data, but in semilogarithmic scale. The results are consistent with a central charge $c =3 \times 0.6528 \approx 2$. $\chi$ is the MPS bond dimension.}
\label{FigScal}
\end{figure}

\section{Effective critical field theory} The transitions between $SO(5)$ Haldane/IH and IH/large-$D$ were studied originally in Ref.~\cite{TuOrus}. There it was shown that the system can be described at low energies using an effective field theory of five Majorana fermions $\xi^{a}$ ($a=1,\ldots ,5$):%
\begin{eqnarray}
\mathcal{H}_{\mathrm{eff}} &=&-iv\sum_{a=1}^{5}(\xi _{R}^{a}\partial _{x}\xi
_{R}^{a}-\xi _{L}^{a}\partial _{x}\xi _{L}^{a})-im_{1}\sum_{a=1}^{2}\xi
_{R}^{a}\xi _{L}^{a}  \notag \\
&&-im_{2}\sum_{a=3}^{4}\xi _{R}^{a}\xi _{L}^{a}-im_{3}\xi _{R}^{5}\xi
_{L}^{5},  \label{eq:FT}
\end{eqnarray}%
where $v$ and $m_{a}$ are velocity and masses of the Majoranas. For the IH (or $T_0$) phase, three fermions have negative masses $m_2$ and $m_3$, and two have positive mass $m_1$, producing three Majorana edge modes forming spin-$1/2$ edge states \cite{lech}.  The phase transitions happen whenever $m_1 = 0$ ($SO(5)$/IH), or $m_2 = 0$ (IH/large-$D$). In both cases two Majorana fermions become massless, and hence these transitions correspond to a CFT with central charge $c = 2 \times \frac{1}{2} = 1$.

This effective field theory treatment may also help to understand the transition between the different effective spin-1 SPT phases. From the numerics, we see that  $T_x$, $T_y$ and $T_z$ also have spin-$1/2$ edge modes because of the two-fold degeneracy in the entanglement spectrum. If we try to describe them using the Majorana field theory, then three Majorana masses must be negative and two positive. If this is the case, then the only possibility is that the three Majoranas with negative mass are \emph{different} from those in the $T_0$ phase (otherwise the phases could be deformed into each other without closing the gap). Thus, if the field theory is correct in this regime, then phase transitions amongst the $T_0, T_x, T_y$ and $T_z$ phases can be viewed as processes of \emph{sign exchange amongst Majorana masses}. To change their sign, some of these masses must be zero at some point, which implies a quantum phase transition described by a CFT. To check the plausible validity of this picture we compute the central charge of the $T_y-T_0$ transition in Fig.~\ref{FigJ3} using the scaling of the entanglement entropy of a block of length $L$. For a CFT such a scaling obeys $S(L) \sim \frac{c}{3}\log L + O(1/L)$ for $L \gg 1$ \cite{entropy}, and hence we can extract the central charge $c$ from an appropriate fit to the data. Our result in Fig.~\ref{FigScal} agrees with $c =3 \times 0.6528 \approx 2$ \footnote{Other $c \approx 2$ quantum critical points have been found in spin-2 chains, see, e.g., Refs.~\cite{pochung,c2}.}, which means that 4 out of the 5 Majoranas would become massless at criticality and interchange the signs of their masses. A similar behaviour is found for the other transitions between the effective spin-1 SPT phases in Figs.~\ref{FigJ1},\ref{FigJ3},\ref{FigJ4}.
 
\section{Other features} The main purpose of this paper is the study of the topologically-nontrivial phases emerging from the model in Eq.~(\ref{eq:Hamiltonian}). Yet, a number of other features appear in the phase diagram, which we now discuss very briefly. For instance, we find a candidate for a gapless phase $gJ_4$ with entanglement spectrum degeneracy of 2, no string order, and topologically trivial. We conjecture this phase to be in the $XY$ universality class, as also found in other spin-2 chains \cite{IHall}. We also find several large- and low-$J$ phases at every plane which are topologically trivial, and which correspond to different symmetry-breaking orders (e.g., ferromagnetic). An in-depth analysis of all these phases, together with the phase diagram of the model in the full parameter space, will be considered in a future study.

\section{Conclusions} Here we have studied a spin-2 model exhibiting a wide variety of SPT phases protected by $(\mathbb{Z}_2 \times \mathbb{Z}_2) + T$ symmetry. In particular, we have found \emph{four} different intermediate effective spin-1 SPT phases, with \emph{continuous} phase transitions between them corresponding to a $c=2$ CFT. Indirect evidence suggests the possibility of an effective field theory of four massless Majorana fermions for such CFT. Our results show, for the first time, that different spin-1 SPT phases separated by quantum critical points can emerge from a single, quite simple, quantum spin chain.

\acknowledgements
We acknowledge F. Pollmann for crucially suggesting the implementation of the techniques in Ref.~\cite{param}. Discussions with N. Bl\"umer, Z.-X. Liu, L. Mazza, T. Nishino, M. Oshikawa, and M. Rizzi are also acknowledged.  A.K. and R.O. acknowledge funding from the JGU and the DFG. H.H.T. acknowledges funding from the EU project SIQS.

\vspace{10pt}
\emph{Note added:}Ê after completion of this paper, another work appeared dealing with phase transitions between SPT phases, see Ref.\cite{note}.

\appendix

\section{The 16 SPT phases in $1d$ protected by $(\mathbb{Z}_2 \times \mathbb{Z}_2) + T$ symmetry}

In Ref.~\cite{wen2} a complete classification of all SPT phases protected by $(\mathbb{Z}_2 \times \mathbb{Z}_2) + T$ symmetry is provided, based on the grounds of group-theory properties. This classification can be summarized in Table I of that paper, where it is seen that there are a total of 16 different phases, and which introduces the notation $T_0, T_x, T_y$ and $T_z$ used in this paper for the typical spin-1 phases.

However, it is possible to understand in a much simpler way the 16 phases in spin chains with $(\mathbb{Z}_2 \times \mathbb{Z}_2) + T$ symmetry. This goes as follows: in MPS language, we have a set of (projective) symmetry operators $R_x,R_z$ and $R_t$ acting on the bond indices. Without loss of generality, we choose $R_x^2 = R_z^2 = \mathbb{I}$ (if $R_x^2 = -\mathbb{I}$, it is always possible to redefine $R_x$ and $R_z$ by multiplying a factor $i$, so $R_x^2 = \mathbb{I}$ is just a gauge choice). The nontrivial sign $\omega = \pm 1$ denotes the commutation relation between $R_x$ and $R_z$, which cannot be gauged away.
This means, $R_x R_z = \omega R_z R_x$. Together with the sign $\beta = \pm 1$ defined from $R_t^2 = \beta \mathbb{I}$, there are in total four choices:

\begin{enumerate}[1)]
\item{$\omega = 1, \beta = 1$}
\item{$\omega = 1, \beta = -1$}
\item{$\omega = -1, \beta = 1$}
\item{$\omega = -1, \beta = -1$}
\end{enumerate}

For each of the above four choices, there are still four possibilities:

\begin{enumerate}[a)]
\item{$[R_z, R_t] = 0$ and $[R_x, R_t] = 0$.}
\item{$[R_z, R_t] = 0$ and $\{R_x, R_t\} = 0$.}
\item{$\{R_z, R_t\} = 0$ and $[R_x, R_t] = 0$.}
\item{$\{R_z, R_t\} = 0$ and $\{R_x, R_t\} = 0$.}
\end{enumerate}

Other commutators are all fixed by the above relations. The above four choices correspond to the index $\gamma$ in Table I of Ref.~\cite{wen22}, and we label them by the values of signs $\mu = \pm 1$ and $\nu = \pm 1$ defined respectively from $R_x R_t = \mu R_t R_x$ and $R_z R_t = \nu R_t R_z$. Thus, there are Êin totalÊ $4 \times 4 = 16$ phases (1a, 1b, 1c, 1d, 2a, 2b,..., 4d), which can be labeled uniquely by the four signs $(\beta, \omega, \mu, \nu)$.

Let us mention that in Ref.~\cite{wen2} operators $R_x, R_z$ and $R_t$ do not correspond to the gauge that we are using here, where $R_x^2 = R_z^2 = \mathbb{I}$. To recover the above results, one has to remove a factor ``i" for $R_x$ and $R_z$ in Table I of that reference, so that they square to $\mathbb{I}$. Then one can check that the commutation relations among $R_x$,$R_z$ and $R_t$ just correspond to the above 16 cases.

\section{Extracting $(\beta, \omega, \mu, \nu)$ from a Matrix Product State.}

For an infinite MPS with one-site translation invariance, Ref.~\cite{param} explains in Eqs.10-12 how to obtain explicitly the operators $U_x \equiv R_x, U_z \equiv R_z$ and $U_t \equiv R_t K$. Here $K$ is the complex conjugation operation, defined by $KAK^{-1} = A^*$ and $K =ÊK^{-1}$. The procedure explained in Ref.~\cite{param} is just a simple MPS calculation, and we address the interested reader to that reference for further information.

Once the above matrices have been determined following Ref.~\cite{param}, we normalize them so that $R_x^2 = R_z^2 = \mathbb{I}$, and $R_t^2 = U_t K U_t K = U_t U_t^* = \pm \mathbb{I}$. Let us assume that all these matrices are $\chi \times \chi$, with $\chi$ the MPS bond dimension. It is easy to see that parameters $(\beta, \omega, \mu, \nu)$ can now be computed as

\beqa
\beta &=& \frac{1}{\chi} {\rm tr}{(R_t^2)} = \frac{1}{\chi} {\rm tr}{(U_t K U_t K)} =  \frac{1}{\chi} {\rm tr}{(U_t U_t^*)} \\
\omega &=& \frac{1}{\chi} {\rm tr}{(R_xR_zR_x^{\dagger}R_z^{\dagger})} \\
\mu &=& \frac{1}{\chi} {\rm tr}{(R_xR_tR_x^{\dagger}R_t^{\dagger})} = \frac{1}{\chi} {\rm tr}{(R_xU_tKR_x^{\dagger}KU_t^{\dagger})}  \nonumber \\
&=&  \frac{1}{\chi} {\rm tr}{(R_xU_tR_x^{T}U_t^{\dagger})} \\
\nu &=& \frac{1}{\chi} {\rm tr}{(R_zR_tR_z^{\dagger}R_t^{\dagger})} = \frac{1}{\chi} {\rm tr}{(R_zU_tKR_z^{\dagger}KU_t^{\dagger})} \nonumber \\ 
&=&  \frac{1}{\chi} {\rm tr}{(R_zU_tR_z^{T}U_t^{\dagger})} .
\eeqa

This procedure can be generalized very easily to the case of, e.g., two-site translation invariance, as is the case of the infinite MPS produced with the standard infinite Time Evolving Block Decimation method \cite{itebd}. In this way, we are able to determine precisely to which one of the 16 SPT phases protected by $(\mathbb{Z}_2 \times \mathbb{Z}_2) + T$ symmetry belongs for a given MPS.

\end{document}